\documentclass[12pt]{iopart}

\usepackage{iopams}
\usepackage{cite}

\begin{document}

\title[Regularization of the big bang singularity]{Regularization of the big bang singularity with a time varying equation of state $w > 1$}

\author{BingKan Xue$^1$\footnote{Present address:\ School\ of\ Natural\ Sciences,\ Institute\ for\ Advanced\ Study,\ Princeton,\ NJ\ 08540,\ USA}, Edward Belbruno$^2$}

\address{$^1$ Department of Physics, $^2$ Department of Astrophysical Sciences,}
\address{Princeton University, Princeton, NJ 08544, USA}

\eads{\mailto{bxue@princeton.edu}, \mailto{belbruno@princeton.edu}}

\begin{abstract}
We study the classical dynamics of the universe undergoing a transition from contraction to expansion through a big bang singularity. The dynamics is described by a system of differential equations for a set of physical quantities, such as the scale factor $a$, the Hubble parameter $H$, the equation of state parameter $w$, and the density parameter $\Omega$. The solutions of the dynamical system have a singularity at the big bang. We study if the solutions can be regularized at the singularity in the sense of whether they have unique branch extensions through the singularity. In particular, we consider the model in which the contracting universe is dominated by a scalar field with a time varying equation of state $w$, which approaches a constant value $w_c$ near the singularity. We prove that, for $w_c > 1$, the solutions are regularizable only for a discrete set of $w_c$ values that satisfy a coprime number condition. Our result implies that the evolution of a bouncing universe through the big bang singularity does not have a continuous classical limit unless the equation of state is extremely fine-tuned.
\end{abstract}


\submitto{\CQG}

\maketitle

\section{Introduction} \label{sec:intro} 

The big bang is commonly considered as a spacetime singularity at the beginning of the expansion of the universe. Such a singularity exists in many theories of the early universe. In inflationary cosmology, for example, it is proved that the spacetime is past incomplete \cite{Borde:2001nh}, implying that there is a singularity in the finite past. However, this singularity need not be the absolute beginning of the universe. In bouncing cosmology, for example, it is conjectured that there is a contraction phase of the universe before the expansion phase. Under general conditions as considered in the singularity theorem \cite{Hawking:1969sw}, the cosmic contraction would end in a spacetime singularity \footnote{There are, however, \emph{nonsingular} bouncing models in which the universe smoothly transitions from contraction to expansion without passing through a spacetime singularity \cite{Xue:2013}. Such models require a violation of the Null Energy Condition \cite{Hawking:1969sw}, which will not be considered in this paper.}. In that scenario, the big bang is both the end of the contraction phase and the beginning of the expansion phase.

For bouncing cosmology, it is essential to describe the transition of the universe from the contraction phase to the expansion phase. A complete set of physical quantities have to be evolved through the singularity. In principle, near the singularity when the energy density of the universe reaches the Planck scale, quantum gravity effects would become important. Presumably the physical quantities would follow unitary quantum evolution that takes them smoothly through the singularity. From a classical point of view, at length scales much larger than the Planck scale, the physical quantities would effectively vary continuously through the singularity. In practice, various matching conditions have been proposed to connect the values of those physical quantities before and after the singularity \cite{Israel:1966rt, Hwang:1991an, Deruelle:1995kd, Finelli:2001sr, Durrer:2002jn, Bars:2011aa}.

Here we take a different approach and analyze the structure of the singularity from a dynamical system's perspective. Given the dynamics of the physical quantities, what are the general conditions under which the solutions can be extended through the singularity? (The term singularity is defined in Section 3.) This question was first addressed in Reference \cite{Belbruno:2013}, which proves the necessary and sufficient conditions for the solutions of the Friedmann equations to be regularizable at the big bang singularity. The term ``regularizability'' refers to whether the solution of a dynamical system ending in a singularity can be uniquely extended to another solution beginning at the same singularity through analytic continuation \cite{McGehee:1981}. In the work of \cite{Belbruno:2013} it is assumed that the universe is homogeneous and includes energy components representing matter, radiation, cosmological constant, spatial curvature, anisotropy, and an additional component with a constant equation of state $w$, which dominates the contracting universe if $w > 1$. The regularization of the singularity is accomplished by using a McGehee transformation \cite{McGehee:1981}, and requires that $w$ belongs to a \emph{discrete} set of values satisfying a coprime number condition \cite{Belbruno:2013}.

In this paper, we consider a more general case where the equation of state of the dominant energy component is not strictly a constant. In particular, this energy component is modeled as a scalar field with an exponential potential, which is widely used in inflationary and bouncing cosmologies \cite{Lucchin:1984yf, Halliwell:1986ja, Burd:1988ss, Steinhardt:2001st, Lehners:2011kr}. In such models, the equation of state $w$ of the scalar field approaches a constant value $w_c$ during the contraction phase, mimicking a perfect fluid with a constant equation of state equal to $w_c$. However, since $w$ varies \emph{continuously} in this case, the coprime number condition cannot be satisfied at all times. Therefore it is important to study if the solutions of the physical quantities can be regularized at the big bang singularity in this more general case.

We show that such a regularization does exist. In the general case, the dynamical system consists of the Friedmann equations for the Hubble parameter $H$ and the equation of motion for the scalar field, or equivalently its equation of state $w$. Additional physical variables are included to represent the relative density of other energy components, such as matter, radiation, cosmological constant, spatial curvature, and anisotropy. The solutions of the dynamical equations have a singularity which corresponds to the big bang. We prove that there exists a unique extension of the solutions through the singularity, provided that the value of $w_c$, i.e. the limit of $w$ at the singularity, satisfies a coprime number condition. This condition is shown to be identical to that obtained in \cite{Belbruno:2013} for a constant $w$. We note that the discrete set of values that satisfy the coprime number condition has measure zero in $\mathbb{R}$, implying that $w_c$ has to be extremely fine-tuned for the physical variables to have classical solutions extending through the singularity; otherwise the evolution of the physical variables does not have a continuous classical limit and must be treated with quantum gravity.

This paper is organized as follows. Section~\ref{sec:1} presents the system of dynamical equations for a set of physical quantities that we consider. Section~\ref{sec:2} gives the definition of singularity and regularizability, and states the main theorem on the necessary and sufficient condition for the solutions of the dynamical system to be regularizable. The theorem is proved in two steps by using the Stable Manifold Theorem. The physical implications of the result are discussed in Section~\ref{sec:3}.

\section{Dynamical system} \label{sec:1} 

In this section we derive the differential equations for a set of variables that describe the evolution of the universe. The Friedmann equations lead to a differential equation for the Hubble parameter $H$, or its reciprocal, $Q \equiv 1/H$. The equation of motion for the scalar field $\phi$ determines its time varying equation of state $w$. With additional energy components besides the scalar field, we introduce the relative energy density $\Omega_m$ for each component and derive the differential equations for them.

Consider a homogeneous, flat, and isotropic universe with the metric
\begin{equation}
ds^2 = - dt^2 + a(t)^2 |d{\bf x}|^2 ,
\end{equation}
where $t$ is the proper time and $\mathbf{x} = (x^1, x^2, x^3)$ are the spatial coordinates. Here $a(t)$ is the scale factor of the universe, and the Hubble parameter $H$ is given by $H \equiv \dot{a} / a$, where the dot $\dot{}$ denotes the derivative with respect to time $t$. $H$ is negative during cosmic contraction, and positive after the universe transitions to expansion. The big bang is at $a = 0$, which is chosen to correspond to the time $t = 0$.

Assume that the scalar field $\phi$ has the Lagrangian
\begin{equation}
\mathcal{L} = \sqrt{-g} \Big[ -\case{1}{2} (\partial \phi)^2 - V(\phi) \Big] ,
\end{equation}
where the potential $V(\phi)$ is an exponential function $V(\phi) = V_0 \, e^{- c \, \phi}$; it is assumed that $V_0 < 0$ and $c > \sqrt{6}$ for reasons explained below. In the homogeneous case, the energy density and pressure of the scalar field are
\begin{equation}
\rho_\phi = \case{1}{2} \, \dot{\phi}^2 + V(\phi), \quad p_\phi = \case{1}{2} \, \dot{\phi}^2 - V(\phi).
\label{eq:rhoP}
\end{equation}
The equation of state parameter $w$ is given by
\begin{equation}
w = {p_\phi} / {\rho_\phi} .
\label{eq:w}
\end{equation}
For our study, we focus on the case with $w > 1$ and $\rho_\phi > 0$; hence $V(\phi) < 0$, which explains the choice of $V_0 < 0$ \footnote{Such a negative exponential potential is typical of \emph{ekpyrotic} models \cite{Steinhardt:2001st, Lehners:2011kr}.}.
The equation of motion for the $\phi$ field is
\begin{equation}
\ddot{\phi} + 3 H \dot{\phi} + V_{\phi} = 0 ,
\label{eq:phiDE}
\end{equation}
where $V_{\phi} \equiv \partial{V}/\partial{\phi}$. This equation can be equivalently written as
\begin{equation}
\dot{\rho}_\phi + 3 H (1 + w) \rho_\phi = 0 .
\label{eq:rhoDE}
\end{equation}

Consider first the case in which the scalar field $\phi$ is the only energy component in the contracting universe. Then the Hubble parameter $H$ obeys the Friedmann equations,
\numparts
\begin{eqnarray}
H^2 = \case{1}{3} \, \rho_\phi , \label{eq:FriedmannEqus1} \\*[4pt]
\dot{H} = -\case{1}{2} (\rho_\phi + p_\phi) . \label{eq:FriedmannEqus2}
\end{eqnarray}
\endnumparts
Using equation~(\ref{eq:w}), the Friedmann equations yield a differential equation for $H$,
\begin{equation}
\dot{H} = - \case{3}{2} (1 + w) H^2 .
\label{eq:HDiffEqu}
\end{equation}
Since $H$ is negative during contraction, for $w > -1$ (as required by the null energy condition), $H \to -\infty$ at the big bang.

Differentiating (\ref{eq:w}) with respect to $t$ and using (\ref{eq:rhoP}), (\ref{eq:rhoDE}), and (\ref{eq:FriedmannEqus1}, \ref{eq:FriedmannEqus2}), one obtains a differential equation for $w$,
\begin{equation}
\dot{w} = H (w - 1) \Big[ 3 (1 + w) - c \sqrt{3 (1 + w)} \Big] ,
\label{eq:wdot2}
\end{equation}
where we used the relation $V_{\phi} / V = -c$, and assumed $\dot{\phi} < 0$ as verified below. Let
\begin{equation}
w_c \equiv \frac{c^2}{3} - 1 ,
\end{equation}
then equation~(\ref{eq:wdot2}) can be written as
\begin{equation}
\dot{w} = \frac{3H \sqrt{1 + w}}{\sqrt{1 + w_c} + \sqrt{1 + w}} \, (w - 1) (w - w_c) .
\label{eq:wdot}
\end{equation}
It can be seen that, for $c > \sqrt{6}$ and hence $w_c > 1$, $w = w_c$ is a fixed point attractor in a contracting universe where $H < 0$. This is the case that we consider in this paper. Note that the attractor solution is described by $\dot{\phi}= 2 / (c \, t)$, where $t < 0$ during contraction; hence, near the fixed point, $\dot{\phi} < 0$ is satisfied.

If there exist other energy components in the universe in addition to the scalar field $\phi$, then equations~(\ref{eq:HDiffEqu}) and (\ref{eq:wdot}) need to be modified. Consider energy components with constant equations of state $w_m$, where $w_m = 0$, $\frac{1}{3}$, $-1$, $-\frac{1}{3}$, or $1$, if the additional energy components represent matter, radiation, cosmological constant, spatial curvature, or anisotropy, respectively. The energy density $\rho_m$ of each component obeys an equation similar to (\ref{eq:rhoDE}),
\begin{equation}
\dot{\rho}_m + 3 H (1 + w_m) \rho_m = 0 .
\label{eq:rhomDE}
\end{equation}

To describe the extra degrees of freedom, introduce the density parameters $\Omega_m$ defined by
\begin{equation}
\Omega_m \equiv \rho_m / \rho_{\rm tot} , \quad \rho_{\rm tot} = \rho_\phi + \sum_m \rho_m ,
\end{equation}
which represent the fractional density of each energy component. The Friedmann equations~(\ref{eq:FriedmannEqus1}, \ref{eq:FriedmannEqus2}) are modified as
\numparts
\begin{eqnarray}
H^2 = \case{1}{3} \, \rho_{\rm tot} = \case{1}{3} \sum_i \rho_i , \label{eq:FriedmannEqusOne} \\*
\dot{H} = - \case{1}{2} (\rho_{\rm tot} + p_{\rm tot}) = -\case{1}{2} \sum_i \rho_i (1 + w_i) , \label{eq:FriedmannEqusTwo}
\end{eqnarray}
\endnumparts
where in these two equations the summation is over all energy components including the scalar field $\phi$.

For illustration, consider just one additional component besides the scalar field. Let the equation of state and the density parameter of this component be denoted by $w_1$ and $\Omega_1$. The Friedmann equations yield the differential equation
\begin{equation}
\dot{H} = - \case{3}{2} \big[ (1 + w) - (w - w_1) \Omega_1 \big] H^2 .
\label{eq:HDiffEquTwo}
\end{equation}
Similarly, equation~(\ref{eq:wdot2}) for the equation of state $w$ of the scalar field $\phi$ is modified as
\begin{equation}
\dot{w} = H (w - 1) \Big[ 3 (1 + w) - c \sqrt{3 (1 + w) (1 - \Omega_1)} \Big] .
\label{eq:wdot2Two}
\end{equation}
Using (\ref{eq:rhoDE}) and (\ref{eq:rhomDE}), one obtains an additional equation for $\Omega_1$,
\begin{equation}
\dot{\Omega}_1 = 3 H (w - w_1) \Omega_1 (1 - \Omega_1) .
\label{eq:dotOmega}
\end{equation}
We assume that $w_1 \leq 1$ \footnote{This requirement also guarantees that the speed of sound for this component is non-superluminal; indeed, for a constant $w_m \leq 1$, the speed of sound is $c_m^2 \equiv \delta p_m / \delta \rho_m = w_m \leq 1$. Incidentally, for the (canonical) scalar field, although $w(t) > 1$, the speed of sound is always $c_s^2 = 1$ (e.g. \cite{Mukhanov:2005}).}, which is true for any $w_1 \in \{ 0, \frac{1}{3}, -1, -\frac{1}{3}, 1 \}$, corresponding to matter, radiation, cosmological constant, spatial curvature, or anisotropy. From equation (\ref{eq:dotOmega}), it can be seen that, for $H < 0$ and $w > 1$, $\Omega_1$ decreases towards the fixed point $\Omega_1 = 0$. That means the relative density of an additional energy component diminishes during the contraction phase; indeed, comparing equations (\ref{eq:rhoDE}) and (\ref{eq:rhomDE}), one sees that the energy density of the scalar field grows exponentially faster than that of any other energy component.

Equations~(\ref{eq:HDiffEquTwo}, \ref{eq:wdot2Two}, \ref{eq:dotOmega}) have a singularity at the big bang when $t \to 0^-$. Indeed, when $H \to -\infty$, the right hand side of (\ref{eq:HDiffEquTwo}, \ref{eq:wdot2Two}, \ref{eq:dotOmega}) become undefined. Therefore the solutions of $H(t)$, $w(t)$, and $\Omega_1(t)$ need to be regularized in order to extend through the singularity. For that purpose, we introduce the variables $Q \equiv 1/H$ and $W \equiv w - w_c$, which both go to $0$ at the singularity. Then equations~(\ref{eq:HDiffEquTwo}, \ref{eq:wdot2Two}, \ref{eq:dotOmega}) can be written in terms of these new variables as follows.

\medskip

\noindent
\textbf{Summary~1.} \; For a homogeneous universe filled with the scalar field $\phi$ and an additional energy component with a constant equation of state $w_1$, the variables $Q$, $W$, and $\Omega_1$ satisfy a system of dynamical equations
\numparts
\begin{eqnarray}
\hspace{-0.5in} \dot{Q} = \case{3}{2} \Big[ (W + w_c + 1) - (W + w_c - w_1) \Omega_1 \Big] , \label{eq:DiffEqusMoreGeneralH} \\*[4pt]
\hspace{-0.5in} \dot{W} = \case{3 \sqrt{W + w_c + 1} \, / Q}{\sqrt{W + w_c + 1} + \sqrt{(w_c + 1) (1 - \Omega_1)}} \, (W + w_c - 1) \big( W + (1 + w_c) \Omega_1 \big) , \label{eq:DiffEqusMoreGeneralW} \\*[4pt]
\hspace{-0.5in} \dot{\Omega}_1 = \case{3 (W + w_c - w_1)}{Q} \, \Omega_1 (1 - \Omega_1) . \label{eq:DiffEqusMoreGeneralOmega}
\end{eqnarray}
\endnumparts

\medskip

\noindent
\textbf{Remark.} \; The above case can be generalized to include more energy components with constant equations of state $w_m$, $m = 1, 2, \cdots$. The generalized differential equations for $Q$, $W$, and $\Omega_m$ are given by
\numparts
\begin{eqnarray}
\hspace{-1in} \dot{Q} = \case{3}{2} \bigg[ (W + w_c + 1) - \sum_m (W + w_c - w_m) \Omega_m \bigg] , \label{eq:DiffEqusMostGeneralH} \\*[4pt]
\hspace{-1in} \dot{W} = \case{3 \sqrt{W + w_c + 1} \, / Q}{\sqrt{W + w_c + 1} + \sqrt{(w_c + 1) (1 - \sum_m \Omega_m)}} \, (W + w_c - 1) \Big( W + (1 + w_c) \sum_m \Omega_m \Big) , \label{eq:DiffEqusMostGeneralW} \\*[4pt]
\hspace{-1in} \dot{\Omega}_m = \case{3}{Q} \, \Omega_m \bigg[ (W + w_c - w_m) (1 - \Omega_m) - \sum_{n \neq m} (W + w_c - w_n) \Omega_n \bigg] , \quad m = 1, 2, \cdots \label{eq:DifEqusMostGeneralOmega}
\end{eqnarray}
\endnumparts
where the summation is over the additional energy components besides the scalar field.

\medskip

Below we consider the case with just two components, the scalar field $\phi$ and one additional energy component with a constant equation of state $w_1$, as presented in Summary~1. For clarity, we denote the density parameter $\Omega_1$ of the additional component simply by $\Omega$. We show that the solutions of $Q(t)$, $W(t)$, $\Omega(t)$, as well as $a(t)$, can be uniquely extended through the big bang singularity at $t = 0$, provided that the value of $w_c$ belongs to a set of rational numbers satisfying a coprime number condition. Our results can be easily generalized to the case with more energy components.

\section{Regularization} \label{sec:2}

In this section we consider the solutions $a(t)$, $Q(t)$, $W(t)$, and $\Omega(t)$ for $t < 0$, which tend to $a = 0$, $Q = 0$, $W = 0$, and $\Omega = 0$ when $t \to 0^-$. We determine necessary and sufficient conditions for these solutions to have a well defined unique extension to $t \geq 0$.

We use regularization methods that are traditionally used in classical and celestial mechanics. There are different types of regularizations \cite{Belbruno:2004}, which all require a change of variables as well as a time transformation. The strongest type of regularization, {\it global regularization}, transforms the singularity into a regular point at which the transformed differential equations become well defined. The transformed solutions can be extended through the singularity in a real analytic manner. One such example is the collision of two point masses in the Newtonian two-body problem, which can be globally regularized to describe a smooth bounce by using the Levi-Civita transformation. Another type of regularization, {\it branch regularization}, reduces the singularity to a fixed point using a transformed time variable. Accordingly, the trajectory of an individual solution flowing to the fixed point can be uniquely matched to the trajectory of another solution emerging from the same fixed point. For such a regularization, the solutions in the original variables and time can be extended through the singularity in a continuous manner. There are also singularities that cannot be regularized at all. In such cases, not a single solution can be extended through the singularity. This occurs, for example, in a triple collision in the Newtonian three-body problem.

For our problem, the solutions to the dynamical system can be branch regularized. The term ``branch regularization'' is formally defined as in Reference~\cite{McGehee:1981}. Consider a system of first order differential equations,
\begin{equation}
\mathbf{X}' = \mathbf{F}(\mathbf{X}) ,
\label{eq:Gen}
\end{equation}  
where $\mathbf{X} \in \mathbb{R}^n$, and $\mathbf{F}$ is a real analytic vector field on an open set $U \subset \mathbb{R}^n$. Here the prime ${}^\prime$ denotes the derivative with respect to a time variable $\tau$. For an initial condition $\mathbf{X}(\tau_0) \in U$, there is a unique solution $\mathbf{X}(\tau)$ that is a real analytic function of $\tau$. Assume that this solution can be maximally extended to an interval $\tau_1 < \tau < \tau_2$, where $-\infty \leq \tau_1 < \tau_0 < \tau_2 \leq \infty$. Then the terms ``singularity'', ``branch extension'', and ``branch regularization'' can be defined as follows.

\medskip

\noindent
\textbf{Definition~1.} \; If $ -\infty < \tau_1$, then the solution is said to begin at $\tau = \tau_1$; and if $\tau_2 < \infty$, then the solution is said to end at $\tau = \tau_2$. In either case, $\tau_1$ or $\tau_2$ is said to be a \emph{singularity} of the solution $\mathbf{X}(\tau)$.

\medskip

It is noted that the term singularity can be used in different ways, such as in dynamical systems when a vector field vanishes, or in General Relativity when some of the curvature invariants diverge. In this paper, we use it as in Definition 1, denoting a point in time, usually an endpoint of a maximal time interval of existence of a solution. This occurs, in this paper, when a time derivative is undefined, e.g. when $\mathbf{X}' \to \infty$. This definition of singularity is commonly used in celestial mechanics (see Reference~\cite{Belbruno:2004}).

\medskip

\noindent
\textbf{Definition~2.} \; Suppose that $\mathbf{X}_1(\tau)$, $\mathbf{X}_2(\tau)$ are two solutions of (\ref{eq:Gen}), and that $\mathbf{X}_1$ ends in a singularity at time $\tau^*$ and $\mathbf{X}_2$ begins in a singularity at the same time. If there exists a multivalued analytic function $\mathbf{X}(\tau)$ having a branch at $\tau^*$ and extending both $\mathbf{X}_1$ and $\mathbf{X}_2$ (i.e. $\mathbf{X}$ coincides with $\mathbf{X}_1$ for $\tau < \tau^*$ and with $\mathbf{X}_2$ for $\tau > \tau^*$), then $\mathbf{X}_2$ is a \emph{branch extension} of $\mathbf{X}_1$ at $\tau^*$, and vice versa.

\medskip

Note that in this Definition, $\mathbf{X}_1$ and $\mathbf{X}_2$ are real analytic functions of real $\tau$, but they are extensions of each other through complex values of $\tau$. In general, the branch extension need not be unique. However, if it is, then we have the following definition.

\medskip

\noindent
\textbf{Definition~3.} \; A solution $\mathbf{X}_1(\tau)$ of (\ref{eq:Gen}) which either begins or ends in a singularity at time $\tau^*$ is said to be \emph{branch regularizable} at $\tau^*$ if it has a unique branch extension $\mathbf{X}_2(\tau)$ at $\tau^*$ and $\mathbf{X}_1(\tau^*) = \mathbf{X}_2(\tau^*)$.

\medskip

Having defined the terms, let us state the main result of the paper. Let $t_0 < 0$ be a certain time in the contracting phase of the universe prior to the big bang at $t = 0$. Consider the system of differential equations given by (\ref{eq:DiffEqusMoreGeneralH}, \ref{eq:DiffEqusMoreGeneralW}, \ref{eq:DiffEqusMoreGeneralOmega}) for $Q(t), W(t), \Omega(t)$ with initial conditions $Q(t_0) < 0$, $W(t_0) + w_c > 1$, and $0 < \Omega(t_0) < 1$. We have the following theorem.

\medskip

\noindent
\textbf{Theorem~1.} \; The solutions $Q(t), W(t), \Omega(t)$ of the dynamical system (18), as well as $a(t)$, are branch regularizable at the singularity $t = 0$ if and only if the value of $w_c$ belongs to a discrete set $\mathbb{Q}_w$ given by
\begin{equation}
\mathbb{Q}_w = \Big\{ w_c =  \case{2 q}{3 p} - 1 \; \Big| \; p, q \in \mathbb{Z}^+ , \; p < q, \; p \perp q, \; q \mbox{ odd} \Big\} .
\end{equation}

\medskip

\noindent
\textbf{Remark.} \; For solutions with constant $w(t) = w_c$, or $W(t) = 0$, Theorem~1 yields the same condition as in Reference~\cite{Belbruno:2013}, where the same set $\mathbb{Q}_w$ is required.

\medskip

Before proving Theorem~1, let us describe the general method we use for the regularization. We first transform $t$ to a new time variable $N$, whereby $t \to 0^-$ corresponds to $N \to \infty$. The singularity at $t = 0$ becomes a fixed point for the system of first order differential equations for $Q$, $W$, $\Omega$ with respect to the new time variable $N$. We show that, for $w_c > 1$, the linearized system of differential equations near the fixed point have purely negative eigenvalues. Therefore the fixed point is an \emph{attractor}, and the solutions near the fixed point all tend to this point exponentially as $N \to \infty$.

The eigenvectors of the linearized system with negative eigenvalues span a stable subspace in which all points flow towards the fixed point. This subspace has a manifold structure and is referred to as a \emph{stable manifold}; we consider this stable manifold in a small neighborhood of the fixed point. Correspondingly, for the nonlinear system, the \emph{local stable manifold} is defined as the set of points in a neighborhood of the fixed point that flow to the fixed point as $N \to \infty$ \cite{Guckenheimer:2002}. In our case, since all eigenvalues are negative, the stable manifold spans the full volume of a neighborhood of the vector space $(Q, W, \Omega)$ near the fixed point $(0, 0, 0)$.

The \emph{Stable Manifold Theorem} states that, under conditions specified below, there exists a homeomorphism (i.e. a one-to-one and continuous mapping) between the stable manifold of the linearized system and the stable manifold of the nonlinear system in a neighborhood of the fixed point. Therefore, solutions of the linear system can be mapped continuously to that of the nonlinear system \cite{Guckenheimer:2002}. More precisely, if the solution of the linearized system is given by $\bar{\mathbf{X}}(\tau)$ near the fixed point, then the solution of the full nonlinear system is given by $\mathbf{X}(\tau) = \mathbf{h}(\bar{\mathbf{X}}(\tau))$, where $\mathbf{h}$ is a homeomorphism.

More explicitly, let the first-order linearized system of (\ref{eq:Gen}) be given by
\begin{equation}
\bar{\mathbf{X}}' = \mathbf{A} \bar{\mathbf{X}} .
\label{eq:Xlin}
\end{equation}
Assume that $\mathbf{A}$ is a constant real matrix and that all the eigenvalues of $\mathbf{A}$ are purely real and non-zero (One can also generalize this to where the eigenvalues have non-zero real part). Therefore, the point $\bar{\mathbf{X}} = \mathbf{0}$ is a hyperbolic fixed point. The full system (\ref{eq:Gen}) can be written as
\begin{equation}
\mathbf{X}' = \mathbf{A} \mathbf{X} + \mathbf{g}(\mathbf{X}) .
\label{eq:Xfull}
\end{equation}
If $\mathbf{g}(\mathbf{X})$ is at least $C^1$, $\mathbf{g}(\mathbf{0}) = \mathbf{0}$, and $\frac{\partial \mathbf{g}}{\partial \mathbf{X}} (\mathbf{0}) = \mathbf{0}$ ($\frac{\partial \mathbf{g}}{\partial \mathbf{X}}$ is a matrix), then the Stable Manifold Theorem states that there exists a homeomorphism $\mathbf{h}$ between the solutions of the linear system (\ref{eq:Xlin}) and that of the full system (\ref{eq:Xfull}) in a neighborhood of $\mathbf{X} = \mathbf{0}$. This theorem requires that $\mathbf{g}(\mathbf{X})$ does not depend on time explicitly.

Note that the version of the Stable Manifold Theorem used here is the Hartman-Grobman theorem \cite{Guckenheimer:2002}. The homeomorphism $\mathbf{h}$ is applied to the points $\bar{\mathbf{X}}$ in the phase space of the linear system, and maps these points homeomorphically to the points $\mathbf{X}$ in the phase space of the nonlinear system. However, within each phase space, $\mathbf{X}$ or $\bar{\mathbf{X}}$, the corresponding solutions $\mathbf{X}(\tau)$ or $\bar{\mathbf{X}}(\tau)$ can both be real analytic functions of $\tau$. Specifically, let $\mathbf{h}^{-1}$ be the inverse of $\mathbf{h}$, and let $\mathbf{X}_0$ be the initial value of $\mathbf{X}$ at $\tau_0$. If the solution of the nonlinear system is $\mathbf{X}(\tau; \mathbf{X}_0)$, which is real analytic in $\tau$ except at the singularity, then $\mathbf{h}$ is such that $\mathbf{h}^{-1} ( \mathbf{X} (\tau; \mathbf{X_0}) ) = \e ^{\mathbf{A} \tau} \, \mathbf{h}^{-1} (\mathbf{X_0})$ is the solution $\bar{\bf{X}}(\tau)$ of the linear system, which is also real analytic as a function of $\tau$.

\medskip

The proof of Theorem~1 involves two main steps. First we show that the time transformation $N(t)$ can be continuously extended through $t=0$ provided that $w_c \in \mathbb{Q}_w$. This is done by writing a second order differential equation for $N(t)$ and appealing to the proof in Reference~\cite{Belbruno:2013}. Then we show that, under the same condition, the solutions to the linearized system of differential equations for $Q$, $W$, and $\Omega$ can be extended through $t=0$. Therefore, using the homeomorphism granted by the Stable Manifold Theorem, the solutions to the full nonlinear system have a branch regularization at the singularity $t=0$.

\medskip

\noindent
\textit{Proof of Theorem~1.} \; For $t < 0$, consider the time transformation given by
\begin{equation}
\frac{dN}{dt} = - H ,
\label{eq:BasicTimeDE}
\end{equation}
where $H < 0$; or, equivalently,
\begin{equation}
\frac{dt}{dN} = - Q .
\label{eq:BasicTimeDE2}
\end{equation}
Since $H = \dot{a}/a$, equation~(\ref{eq:BasicTimeDE}) implies that the new time variable is simply given by
\begin{equation}
N = - \ln a + \hbox{const} .
\label{eq:BasicTrans}
\end{equation}
In the new time variable $N$, equations~(\ref{eq:DiffEqusMoreGeneralH}, \ref{eq:DiffEqusMoreGeneralW}, \ref{eq:DiffEqusMoreGeneralOmega}) can be written as the following system of differential equations for $Q$, $W$, and $\Omega$,
\numparts
\begin{eqnarray}
\hspace{-0.5in} \frac{dQ}{dN} = - \case{3}{2} \Big[ (W + w_c + 1) - (W + w_c - w_1) \Omega \Big] Q , \label{eq:dQdN} \\*[4pt]
\hspace{-0.5in} \frac{dW}{dN} = \case{- 3 \sqrt{W + w_c + 1}}{\sqrt{W + w_c + 1} + \sqrt{(w_c + 1) (1 - \Omega)}} \, (W + w_c - 1) \big( W + (1 + w_c) \Omega \big) , \label{eq:dwdN} \\*[4pt]
\hspace{-0.5in} \frac{d\Omega}{dN} = - 3 (W + w_c - w_1) \Omega (1 - \Omega) . \label{eq:dOmegadN}
\end{eqnarray}
\endnumparts
The singularity at $t = 0$ is mapped to a fixed point at $(Q, W, \Omega) = (0, 0, 0)$ to which the solutions flow as $N \to \infty$. The solutions to the original dynamical system can be obtained from the solutions to the transformed differential equations~(\ref{eq:dQdN}, \ref{eq:dwdN}, \ref{eq:dOmegadN}) and the time transformation (\ref{eq:BasicTimeDE2}). Note that, for $t > 0$, the time transformation should be modified by changing the sign in (\ref{eq:BasicTimeDE2}), since $H > 0$ after the big bang. Accordingly, equations~(\ref{eq:dQdN}, \ref{eq:dwdN}, \ref{eq:dOmegadN}) also change sign, and their solutions correspond to trajectories that approach the fixed point $(Q, W, \Omega) = (0, 0, 0)$ as $N \to -\infty$. The solution of the dynamical system $(Q(t), W(t), \Omega(t))$ for $t < 0$ can be branch regularized if it has a unique branch extension to a solution for $t > 0$. The regularized solution is continuous at $t = 0$ where $(Q, W, \Omega) = (0, 0, 0)$. \footnote{Note that the original variable $H(t)$ is not defined at $t = 0$; however, its solution for $t < 0$ has a unique branch extension to $t > 0$.}

To illustrate our approach, we first consider solutions on certain submanifolds of the $(Q, W, \Omega)$ space, then generalize to the full space.

\medskip

\noindent
\textbf{Case~1.} \; single component, $w =$ const

\medskip

\noindent
We first consider the simplest case where there is only one energy component with a constant $w$. This case corresponds to solutions with $w(t) = w_c$ and $\Omega(t) = 0$. Such solutions remain on the $Q$-axis at $(W, \Omega) = (0, 0)$, which forms an invariant submanifold. Equations~(\ref{eq:dwdN}, \ref{eq:dOmegadN}) are trivially satisfied, and equation~(\ref{eq:dQdN}) becomes
\begin{equation}
\frac{dQ}{dN} = -\frac{3}{2} (1 + w_c) Q .
\label{eq:QDEWConst}
\end{equation}
This equation is already linear in $Q$. It is clear that $Q=0$ is a stable fixed point as $N \to \infty$, since the eigenvalue $\lambda_1 = -\frac{3}{2} (1 + w_c)$ is negative. Therefore the $Q$-axis is a stable submanifold. Equation~(\ref{eq:QDEWConst}) can be explicitly solved, yielding
\begin{equation}
Q(N) \propto - e^{-\frac{3}{2} (1 + w_c) N} .
\label{eq:SolnQN}
\end{equation}

\noindent
\textit{Step~1.} In order to have a regularization of $Q$ in terms of the time $t$ at $t = 0$, we need an extension of $N(t)$ through $t = 0$, or equivalently an extension of $a(t)$ through $t=0$ due to equation~(\ref{eq:BasicTrans}). In the current case it can be done by simply solving equation~(\ref{eq:BasicTimeDE2}) using (\ref{eq:SolnQN}), which gives
\begin{equation}
N = - \frac{2}{3 (1 + w_c)} \, \ln (-t) + \mbox{const} ,
\label{eq:NofT1}
\end{equation} 
and, hence,
\begin{equation}
a(t) \propto (-t)^{2 / 3(1+w_c)} .
\label{eq:SolnsQaoft}
\end{equation}
It can be seen that $a$ as a function of $t$ has a real branch at $t > 0$ provided that
\begin{equation}
\frac{2}{3 (1 + w_c)} = \frac{p}{q} \, ,
\end{equation}
and $(p,q) \in \{ p, q \in \mathbb{Z}^+ \; | \; p < q, \; p \perp q, \; q \mbox{ odd} \}$. This is exactly the condition $w_c \in \mathbb{Q}_w$ in Theorem~1.
\medskip

\noindent
\textit{Step~2.} To find the solution for the variable $Q$, substitute (\ref{eq:NofT1}) into (\ref{eq:SolnQN}), yielding
\begin{equation}
Q(t) \propto t .
\end{equation}
Clearly $Q$ has a unique extension to $t > 0$. This proves the regularizability of solutions on the submanifold of the $Q$-axis.

It is remarked that the proof for this above case can be deduced from the proof in Reference~\cite{Belbruno:2013}, which considers a more general situation with extra energy components of constant $w_m$, including presureless matter, radiation, cosmological constant, spatial curvature, and anisotropy. The above case can be recovered by letting the density of the other components tend to $0$. (This is not as straightforward as the proof presented above, since one must make a slight modification to the proof \cite{Belbruno:2013} to ensure that the equations are well defined when the anisotropy term $\sigma^2$ tends to $0$.)

\medskip

\noindent
\textbf{Case~2.} \; single component (scalar field), $w = w(t)$

\medskip

\noindent
We now consider a scalar field $\phi$ as the only energy component in the universe. This case corresponds to solutions of the dynamical system (\ref{eq:dQdN}, \ref{eq:dwdN}, \ref{eq:dOmegadN}) with $\Omega(t) = 0$, which lie on the invariant submanifold of the $(Q, W)$-plane at $\Omega=0$. Equation~(\ref{eq:dOmegadN}) is trivially satisfied, whereas (\ref{eq:dQdN}, \ref{eq:dwdN}) can be written in terms of the new time variable $N$ as
\numparts
\begin{eqnarray}
\frac{dQ}{dN} = -\case{3}{2} (W + w_c + 1) Q , \label{eq:Qdenew} \\*[4pt]
\frac{dW}{dN} = \case{-3 \sqrt{W + w_c + 1}}{\sqrt{W + w_c + 1} + \sqrt{w_c + 1}} \, (W + w_c - 1) W . \label{eq:wdenew}
\end{eqnarray}
\endnumparts

Near the fixed point at $(Q, W) = (0, 0)$, the above system of differential equations can be written in terms of the vector $\mathbf{X} \equiv (Q, W)^{\rm T}$ as
\begin{equation}
\frac{d \, \mathbf{X}}{dN} = \mathbf{A} \mathbf{X} + \mathbf{\Delta}(\mathbf{X}) ,
\label{eq:GoodSystem}
\end{equation}
where the matrix $\mathbf{A}$ is given by
\begin{equation}
\mathbf{A} = \left( \begin{array}{cc}
-\case{3}{2}(w_c + 1) & 0 \\[4pt]
0 & -\case{3}{2}(w_c - 1) \end{array} \right) ,
\end{equation}
and the vector field $\mathbf{\Delta}$ is given by
\begin{equation}
\mathbf{\Delta} = \left( \begin{array}{c} -\case{3}{2} Q W \\[4pt]
-\frac{3 \big( 1 + 3 w_c + 2 W + 2 \sqrt{(1 + w_c) (1 + w_c + W)} \big)}{2 \big( \sqrt{1 + w_c} + \sqrt{1 + w_c + W} \big)^2} \, W^2 \end{array} \right) .
\label{eq:FCase2}
\end{equation}
It is clear that $\mathbf{A} \mathbf{X}$ is the linear order term and $\mathbf{\Delta} = \mathcal{O}(|\mathbf{X}|^2)$.

Let $\bar{\mathbf{X}} \equiv (\bar{Q}, \bar{W})^{\rm T}$ be the solution to the linearized equations, satisfying
\begin{equation}
\frac{d \, \bar{\mathbf{X}}}{dN} = \mathbf{A} \bar{\mathbf{X}} .
\label{eq:linsys}
\end{equation}
The eigenvalues of $\mathbf{A}$ are given by $\lambda_1 = -\frac{3}{2}(w_c-1)$ and $\lambda_2 = -\frac{3}{2}(w_c+1)$, which are both negative for $w_c > 1$. Hence the $(Q, W)$-plane is a stable submanifold, and the variables $Q$, $W$ flow towards the fixed point $(0, 0)$. By the Stable Manifold Theorem, in a sufficiently small neighborhood of the fixed point, the solutions of the full system (\ref{eq:GoodSystem}) is homeomorphic to that of the linearized system (\ref{eq:linsys}). The solutions of the linear system are given by combinations of the eigenvectors $(\bar{Q}(N),0)^{\rm T}$ and $(0, \bar{W}(N))^{\rm T}$, where
\begin{equation}
\bar{Q}(N) \propto e^{-\frac{3}{2}(w_c+1)N}, \quad \bar{W}(N) \propto e^{-\frac{3}{2}(w_c-1)N} .
\label{eq:GoodSolutions}
\end{equation}

\medskip

\noindent
\textit{Step~1.} As in Case~1, we first show that the time transformation $N(t)$, or equivalently $a(t)$, can be extended through $t=0$. This can be done by appealing to the proof in Reference~\cite{Belbruno:2013} as follows. Let $g(N) \equiv Q(N) - \bar{Q}(N)$, which is the difference between the full solution and the linearized solution, satisfying
\begin{equation}
\lim_{N \to \infty} \frac{g}{Q} = \lim_{N \to \infty} \frac{g}{\bar{Q}} = 0 .
\label{eq:limitgQ}
\end{equation}
Then, by equation~(\ref{eq:BasicTimeDE2}),
\begin{equation}
\dot{N} = - \frac{1}{Q} = - \frac{1}{\bar{Q}} + \frac{g}{\bar{Q} \big( \bar{Q} + g \big)} = - \frac{1}{\bar{Q}} + \bar{g} ,
\end{equation}
where we defined $\bar{g} \equiv g / \bar{Q} \big( \bar{Q} + g \big)$. Differentiating by $t$ again and using (\ref{eq:GoodSolutions}) gives
\begin{eqnarray}
\frac{\ddot{a}}{a} &= \ddot{N} + \dot{N}^2 
= - \frac{(3 w_c + 1)}{2 \bar{Q}^2} + \frac{(3 w_c - 1) \bar{g}}{2 \bar{Q}} + \frac{\bar{g}'}{\bar{Q}} - \bar{g} \bar{g}' + \bar{g}^2 \nonumber \\*
&= - \frac{(3 w_c + 1)}{2 a^{3 (1 + w_c)}} + \frac{(3 w_c - 1) \bar{g} + 2 \bar{g}'}{2 a^{3 (1 + w_c) / 2}} - \bar{g} \bar{g}' + \bar{g}^2 ,
\label{eq:BigEqu}
\end{eqnarray}
where for simplicity we set $\bar{Q} = a^{3 (1 + w_c) / 2}$ using (\ref{eq:GoodSolutions}) and absorbing the proportionality constant by scaling $a$; $\bar{g}(N)$ is to be expressed as a function of $a$ through (\ref{eq:BasicTrans}). The above equation is of the same form as Equation~(12) in \cite{Belbruno:2013},
\begin{equation}
\ddot{a} = - \frac{(3 w_c + 1)}{2 a^{3 w_c + 2}} - f_3(a) ,
\label{eq:GoodForm}
\end{equation}
where in our case $f_3(a) \equiv \big((3 w_c - 1) \bar{g} + 2 \bar{g}' \big) / 2 a^{3 (1 + w_c) / 2} - \bar{g} \bar{g}' + \bar{g}^2$. It is verified in \ref{sec:f3} that $f_3(a)$ goes to infinity slower than the leading term in (\ref{eq:GoodForm}). Given this condition, following the proof of Theorem 5 in \cite{Belbruno:2013}, it can be concluded that there exists a unique extension of $a(t)$ through $t=0$, provided that $w_c \in \mathbb{Q}_w$. Moreover, the leading order solution to equation~(\ref{eq:GoodForm}) without the $f_3$ term,
\begin{equation}
\bar{a}(t) \propto (-t)^{2 / 3 (w_c + 1)} ,
\label{eq:AofT}
\end{equation}
can be mapped to the full solution of $a(t)$ through a homeomorphism. Therefore (\ref{eq:AofT}) gives a leading order time transformation $\bar{N}(t) = - \ln \bar{a}(t) + {\rm const}$, which can be continously mapped to the full time transformation $N(t)$ near the singularity.

\medskip

\noindent
\textit{Step~2.} \; Using the leading order time transformation from (\ref{eq:AofT}), the linear solutions $\bar{Q}$, $\bar{W}$ in (\ref{eq:GoodSolutions}) can be expressed in terms of the time $t$ as
\begin{equation}
\bar{Q}(\bar{N}(t)) \propto t , \quad \bar{W}(\bar{N}(t)) \propto (-t)^{(w_c - 1) / (w_c + 1)} .
\label{eq:WofT}
\end{equation}
Like in Case~1, $\bar{Q}$ can be readily extended through $t=0$. For $\bar{W}$ to have a unique branch extension, the exponent $\frac{w_c-1}{w_c+1}$ must satisfy the condition
\begin{equation}
\frac{w_c - 1}{w_c + 1} = \frac{\tilde{p}}{\tilde{q}} ,
\end{equation}
where $\tilde{p}, \tilde{q}$ are relatively prime integers, $\tilde{q} > \tilde{p} > 0$, and $\tilde{q}$ is odd. In \ref{sec:prime}, it is shown that this condition is equivalent to the coprime number condition, $w_c \in \mathbb{Q}_w$, in Theorem~1. Therefore, for $w_c \in \mathbb{Q}_w$, both solutions in (\ref{eq:WofT}) can be uniquely extended through $t=0$.

Finally, by the Stable Manifold Theorem, the linearized solutions $\bar{Q}(\bar{N})$, $\bar{W}(\bar{N})$ can be mapped by a homeomorphism to the nonlinear solutions $Q(\bar{N})$, $W(\bar{N})$ in a neighborhood of the singularity. Then, using the homeomorphism in Step~1, the time transformation $\bar{N}(t)$ can be continuously mapped to $N(t)$ near the singularity. Putting these together, the solutions $Q(t)$, $W(t)$ of the full system of differential equations~(\ref{eq:Qdenew}, \ref{eq:wdenew}) can be extended through the singularity at $t=0$.

\medskip

\noindent
\textbf{Case~3.} \; two components (scalar field plus constant $w_1$), $w = w(t)$, $\Omega = \Omega(t)$

\medskip

\noindent
This is the general case considered in Theorem~1. The proof is very similar to that of Case~2. The system of differential equations is given by (\ref{eq:DiffEqusMoreGeneralH}, \ref{eq:DiffEqusMoreGeneralW}, \ref{eq:DiffEqusMoreGeneralOmega}) in Summary~1, which are transformed to equations~(\ref{eq:dQdN}, \ref{eq:dwdN}, \ref{eq:dOmegadN}) with the new time variable $N$. Consider a small neighborhood of the fixed point at $(Q, W, \Omega) = (0, 0, 0)$. Let $\mathbf{X} = (Q, W, \Omega)^{\rm T}$, then $\mathbf{X}$ satisfies an equation similar to (\ref{eq:GoodSystem}), but with the matrix $\mathbf{A}$ given by
\begin{equation}
\mathbf{A} = \left( \begin{array}{ccc}
-\case{3}{2} (w_c + 1) & 0 & 0 \\[4pt]
0 & -\case{3}{2} (w_c - 1) & -\case{3}{2} (w^2_c - 1) \\[4pt]
0 & 0 & -3 (w_c - w_1) \end{array} \right) ,
\end{equation}
and the vector field $\mathbf{\Delta}(\mathbf{X})$ given by
\begin{equation}
\hspace{-1in} \mathbf{\Delta}(\mathbf{X}) = \left( \begin{array}{c} -\case{3}{2} Q \big[ W - (w_c + W - w_1) \Omega \big] \\[4pt]
-\frac{3 \big( 1 + 3 w_c + 2 W + 2 \sqrt{(1 + w_c) (1 + w_c + W) (1 - \Omega)} \big) W + 3 (w_c^2 - 1) \Omega}{2 \big( \sqrt{(1 + w_c) (1 - \Omega)} + \sqrt{1 + w_c + W} \big)^2} \, \big( W + (1 + w_c) \Omega \big) \\[4pt]
-3 W \Omega + 3 (w_c - w_1) \Omega^2 + 3 W \Omega^2 \end{array} \right) ,
\label{eq:FCase3}
\end{equation}
from which it is clear that $\mathbf{\Delta} = \mathcal{O}(|\mathbf{X}|^2)$. The eigenvalues of $\mathbf{A}$ are given by the same $\lambda_1, \lambda_2$ as in Case~2, plus $\lambda_3 = -3 (w_c - w_1)$. They are all negative since, by assumption, $w_c > 1 \geq w_1$. Therefore, by the Stable Manifold Theorem, the solutions of the full system of differential equations with $\mathbf{\Delta} \neq \mathbf{0}$ are homeomorphic to the solutions of the linearized system with $\mathbf{\Delta} = \mathbf{0}$ near the fixed point. The solutions of the linear system are given by the eigenvectors $(\bar{Q}(N), 0, 0)^{\rm T}$, $(0, \bar{W}(N), 0)^{\rm T}$, and $(0, 0, \bar{\Omega}(N))^{\rm T}$, where
\begin{equation}
\bar{Q}(N) \propto - e^{-\frac{3}{2}(w_c+1)N} , \quad \bar{W}(N) \propto e^{-\frac{3}{2}(w_c-1)N} , \quad \bar{\Omega}(N) \propto e^{-3(w_c-w_1)N} .
\label{eq:leading}
\end{equation}

\medskip

\noindent
\textit{Step~1.} \; As in Case~2, we first look for the extension of $a(t)$, and hence $N(t)$, through $t=0$. The same differential equation~(\ref{eq:GoodForm}) can be derived for $a(t)$, where in the expression for $f_3(a)$, the function $g \equiv Q - \bar{Q}$ is given by the difference of the full and the linear solutions $Q$ and $\bar{Q}$ in the current case. The same proof as presented in \ref{sec:f3} follows and ensures that $a(t)$ can be extended through $t=0$ provided that $w_c \in \mathbb{Q}_w$. The leading order solution $\bar{a}(t)$ is given by the same equation~(\ref{eq:AofT}), which yields $\bar{N}(t)$ that is homeomorphic to the time transformation $N(t)$.

\medskip

\noindent
\textit{Step~2.} \; Using (\ref{eq:AofT}), the leading order solutions for $\bar{Q}$, $\bar{W}$, and $\bar{\Omega}$ become
\begin{equation}
\bar{Q}(\bar{N}(t)) \propto t , \quad \bar{W}(\bar{N}(t)) \propto (-t)^{\frac{w_c-1}{w_c+1}} , \quad \bar{\Omega}(\bar{N}(t)) \propto (-t)^{\frac{2(w_c-w_1)}{w_c+1}} .
\label{eq:leading_t}
\end{equation}
The first two solutions $\bar{Q}(\bar{N}(t))$ and $\bar{W}(\bar{N}(t))$ are the same as in (\ref{eq:WofT}), requiring the same condition, $w_c \in \mathbb{Q}_w$, for them to be extended through $t=0$. For the value of $w_1$ chosen from $\{ 0, \frac{1}{3}, -1, -\frac{1}{3}, 1 \}$, it can be shown that the same condition $w_c \in \mathbb{Q}_w$ also ensures a unique branch extension of $\bar{\Omega}(\bar{N}(t))$ through $t=0$. Indeed, the exponent in the solution $\bar{\Omega}(\bar{N}(t))$ can be written as
\begin{equation}
\frac{2(w_c-w_1)}{w_c+1} = (3 - 3 w_1) \, \frac{2}{3(w_c+1)} + 2 \, \Big( \frac{w_c-1}{w_c+1} \Big) .
\end{equation}
Since $(3 - 3 w_1)$ is an integer for any $w_1 \in \{ 0, \frac{1}{3}, -1, -\frac{1}{3}, 1 \}$, and both factors $\frac{2}{3(w_c+1)}$ and $\frac{w_c-1}{w_c+1}$ allow branch extensions under the condition $w_c \in \mathbb{Q}_w$ (see \ref{sec:prime}), the above linear combination also guarantees a branch regularizable solution.

Finally, using the Stable Manifold Theorem, the linear solutions $\bar{Q}(\bar{N})$, $\bar{W}(\bar{N})$, $\bar{\Omega}(\bar{N})$ can be homeomorphically mapped to the full solutions $Q(\bar{N})$, $W(\bar{N})$, $\Omega(\bar{N})$ of the dynamical system in a small neighborhood of the fixed point $(Q, W, \Omega) = (0, 0, 0)$. Moreover, the leading order solution $\bar{a}(t)$, as well as $\bar{N}(t)$, can be homeomorphically mapped to the full solutions $a(t)$ and $N(t)$ near the singularity at $t = 0$. Therefore, the solutions $Q(t)$, $W(t)$, $\Omega(t)$, and $a(t)$ can all be uniquely extended through the singularity. This completes the proof of Theorem~1. \hfill $\square$

\section{Conclusion} \label{sec:3}

We have proved that the solutions to the dynamical system given by the Friedmann equations and the equations for the scalar field and other energy components can be regularized at the big bang singularity if and only if the asymptotic value of the equation of state, $w_c$, belongs to a discrete set of values satisfying the coprime number condition. Since our proof only uses the leading order solution of the scalar field equation of state, it may be generalized to other scalar field models in which the equation of state approaches a constant value $w_c$. So far we have considered the evolution of a homogeneous universe described by the scale factor $a(t)$ as the only dynamical variable in the spacetime metric. Incorporating inhomogeneous perturbations in the metric requires adding more variables and differential equations to the dynamical system, which is expected to make the dynamics only more complicated and the regularizability condition more stringent.

The conclusion that only a discrete set of $w_c$ values, of measure zero in $\mathbb{R}$, would permit regularizable solutions is quite striking. It implies that the equation of state of the universe must be extremely fine-tuned in order to pass through the singularity. On the other hand, it could be possible that a quantum resolution of the singularity changes the coprime number condition; then it means that the transition of the universe from contraction to expansion depends crucially on the quantum process and does not have a simple classical limit, which would be an equally surprising result.

In our study we chose the value of $w_c$ to be greater than $1$, so that the scalar field becomes the dominant energy component in the contracting universe, which ensures the attractor solution with $w \to w_c$ as $t \to 0^-$. Such a component also prevents the growth of spatial curvature and anisotropy that can lead to chaotic mixmaster behavior in the contraction phase \cite{Erickson:2003zm, Garfinkle:2008ei}. However, if the attractor solution is to hold all the way up to $t \to 0^-$, then the scalar field potential has to remain exponential as $\phi \to -\infty$, which is clearly unbounded below. In a more realistic setup, e.g. the cyclic model \cite{Steinhardt:2001st}, the potential is assumed to vanish after a certain point before the big bang, whereby the scalar field equation of state approaches $1$. Applying our analysis to this case, it turns out that the system of dynamical equations is no longer linear but at least quadratic at leading order. As a result, the Stable Manifold Theorem cannot be applied. Other methods need to be used to study this case, which we leave for future work.

\ack

We thank David Spergel, Paul Steinhardt, and Frans Pretorius for helpful discussions.

\appendix

\section*{Appendix}
\setcounter{section}{1}

\subsection{Proof that $f_3(a) \to \infty$ at a subdominant rate} \label{sec:f3}

Equation~(\ref{eq:GoodForm}) can be written as
\begin{equation}
\ddot{a} = L(a) - f_3(a) ,
\label{eq:First}
\end{equation}
where $L(a)$ is the leading term,
\begin{equation}
\frac{L(a)}{a} = - \frac{3 w_c + 1}{2 \bar{Q}^2} ,
\end{equation}
and $f_3(a)$ is given by
\begin{equation}
\frac{f_3(a)}{a} = - \frac{(3 w_c - 1) \bar{g}}{2 \bar{Q}} - \frac{\bar{g}'}{\bar{Q}} + \bar{g} \bar{g}' - \bar{g}^2 ,
\end{equation}
where $\bar{g} = - g / \bar{Q} (\bar{Q} - g)$, and $g = \bar{Q} - Q$.
We want to show that

\medskip

\noindent 
\textbf{Proposition~1.}
\begin{equation}
\lim_{a \rightarrow 0} \frac{f_3}{L} = 0 .
\label{eq:Limit}
\end{equation}

\medskip

\noindent
\textit{Proof.} \; Since
\begin{equation}
\hspace{-0.5in} \frac{f_3}{L} = \frac{3 w_c -1}{3 w_c + 1} (\bar{g} \bar{Q}) + \frac{2}{3 w_c + 1} (\bar{g}' \bar{Q}) - \frac{2}{3 w_c + 1} (\bar{g} \bar{Q}) (\bar{g}' \bar{Q}) + \frac{2}{3 w_c + 1} (\bar{g} \bar{Q})^2 ,
\end{equation}
It is clear that the limit (\ref{eq:Limit}) is true if $(\bar{g}\bar{Q}) \rightarrow 0$ and $(\bar{g}'\bar{Q}) \rightarrow 0$ as $a \rightarrow 0$. These two limits are proved in the following two lemmas.

\medskip

\noindent
\textbf{Lemma~1.}
\begin{equation}
\lim_{a \rightarrow 0} (\bar{g}\bar{Q}) = 0 .
\end{equation}

\medskip

\noindent
\textit{Proof.} \; Using the expression $\bar{g} = - g / \bar{Q} (\bar{Q} - g)$, one finds
\begin{equation}
(\bar{g}\bar{Q}) = \frac{- g}{\bar{Q} - g} .
\end{equation}
But, by equation~(\ref{eq:limitgQ}), $g / \bar{Q} \to 0$ as $a \to 0$, hence $(\bar{g}\bar{Q}) \to 0$. \hfill $\square$

\medskip

\noindent 
\textbf{Lemma~2.}
\begin{equation}
\lim_{a \rightarrow 0} (\bar{g}' \bar{Q}) = 0 .
\end{equation}

\medskip

\noindent
\textit{Proof.} \; Directly differentiating $\bar{g}$ gives
\begin{equation}
\bar{g}' = \frac{- \bar{Q}^2 g' + 2 \bar{Q} \bar{Q}' g - \bar{Q}' g^2}{\bar{Q}^2 (\bar{Q} - g)^2} .
\end{equation}
Therefore, using $\bar{Q}' = - \frac{3}{2}(1 + w_c) \bar{Q}$, one finds
\begin{equation}
(\bar{g}' \bar{Q}) = \frac{- 2 \bar{Q} g' - 6 (1 + w_c) \bar{Q} g + 3 (1 + w_c) g^2}{2 (\bar{Q} - g)^2} .
\end{equation}
This goes to zero if, as $a \to 0$, $g / \bar{Q} \to 0$ and $g' / \bar{Q} \to 0$. The former is true by equation~(\ref{eq:limitgQ}), and the latter is proved as follows. Taking the derivative of $g = \bar{Q} - Q$ using equations~(\ref{eq:dQdN}) and (\ref{eq:leading}), one finds
\begin{equation}
\hspace{-0.5in} \frac{g'}{\bar{Q}} = \case{3}{2} \Big[ W - (W + w_c - w_1) \Omega - \big( (W + w_c + 1) (1 - \Omega) + 2 (1 + w_1) \Omega \big) (g / \bar{Q}) \Big] .
\end{equation}
Since $W, \Omega \to 0$ and $g / \bar{Q} \to 0$ as $a \to 0$, $g' / \bar{Q} \to 0$ as well. \hfill $\square$

\medskip

This completes the proof of Proposition~1. \hfill $\square$

\subsection{Equivalence of coprime number conditions} \label{sec:prime}

\noindent
\textbf{Proposition~2.} \; The coprime number condition in Theorem~1 (referred to as \textit{Condition~1}) is equivalent to the coprime number condition in Case~2 (\textit{Condition~2}).

\medskip

\noindent
\textit{Proof.} \; Recall that Condition~1 in Theorem~1 is given by $w_c \in \mathbb{Q}_w$, i.e. there exist integers $p, q$ such that
\begin{equation}
\frac{2}{3 (w_c + 1)} = \frac{p}{q} ,
\label{eq:Cold}
\end{equation}
where $p, q$ are relatively prime, $0 < 3p < q$, and $q$ is odd.
Meanwhile, in Case~2 of the Proof, Condition~2 requires that there exist integers $\tilde{p}, \tilde{q}$ such that
\begin{equation}
\frac{w_c - 1}{w_c + 1} = \frac{\tilde{p}}{\tilde{q}} ,
\label{eq:Cnew}
\end{equation}
where $\tilde{p}, \tilde{q}$ are relatively prime, $\tilde{q} > \tilde{p} > 0$, and $\tilde{q}$ is odd.

To prove the equivalence of Conditions~1 and 2, we first show that Condition~2 implies Condition~1. Assuming Condition~2, i.e. equation~(\ref{eq:Cnew}), one can write the left hand side of (\ref{eq:Cold}) as
\begin{equation}
\frac{2}{3 (w_c + 1)} = \frac{\tilde{q}-\tilde{p}}{3\tilde{q}} .
\end{equation}
Let the integers $p, q$ be either
\begin{equation}
\left\{ \begin{array}{l} p = \tilde{q} - \tilde{p} \\[3pt] q = 3 \tilde{q} \end{array} \right.
\quad \mbox{if} \quad 3 \nmid (\tilde{q} - \tilde{p}) ,
\end{equation}
or
\begin{equation}
\left\{ \begin{array}{l} p = \frac{\tilde{p} - \tilde{q}}{3} \\[3pt] q = \tilde{q} \end{array} \right.
\quad \mbox{if} \quad 3 \mid (\tilde{q} - \tilde{p}) .
\end{equation}
For $\tilde{p}, \tilde{q}$ that satisfy Condition~2, it is clear that $p \perp q$ and $q$ is odd in either case above. Moreover, $\tilde{q} > \tilde{p} > 0$ implies $q > 3p > 0$. Therefore Condition~1 is established.

Similarly, Condition~1 also implies Condition~2. Starting from Condition~1, i.e. equation~(\ref{eq:Cold}), one can write
\begin{equation}
\frac{w_c - 1}{w_c + 1} = \frac{q-3p}{q} .
\end{equation}
Let
\begin{equation}
\left\{ \begin{array}{l} \tilde{p} = q - 3p \\[3pt] \tilde{q} = q \end{array} \right.
\quad \mbox{if} \quad 3 \nmid q ,
\end{equation}
or
\begin{equation}
\left\{ \begin{array}{l} \tilde{p} = \frac{q}{3} - p \\[3pt] \tilde{q} = \frac{q}{3} \end{array} \right.
\quad \mbox{if} \quad 3 \mid q .
\end{equation}
For $p, q$ satisfying Condition~1, it is clear that, in either case above, $\tilde{p} \perp \tilde{q}$ and $\tilde{q}$ is odd. Also, $q > 3p > 0$ implies $\tilde{q} > \tilde{p} > 0$. Therefore Condition~2 is satisfied. This establishes the equivalence of Conditions~1 and 2. \hfill $\square$

\end{document}